\documentclass[aps,pra,twocolumn,showpacs,superscriptaddress,groupedaddress]{revtex4}

\usepackage{graphicx}
\usepackage{dcolumn}
\usepackage{bm}

\usepackage{amsmath} 
\usepackage{esint}
\usepackage{amsfonts}
\usepackage{mathrsfs} 
\usepackage{tabularx}  
\usepackage{booktabs}  
\usepackage{multirow}

\usepackage{braket}
\usepackage{color}
\usepackage[colorlinks,
linkcolor=blue,
anchorcolor=blue,
citecolor=blue]{hyperref}

\begin{document}

\title{Nonlinear Quantum Neuron: A Fundamental Building Block for Quantum Neural Networks}

\author{Shilu Yan}
\address{School of Automation Science and Engineering, South China University of Technology, Guangzhou 510640, China}
\author{Hongsheng Qi}
\address{Key Laboratory of Systems and Control, Academy of Mathematics and Systems Science, Chinese Academy of Sciences, Beijing 100190, China}
\address{School of Mathematical Sciences, University of Chinese Academy of Sciences, Beijing 100049, China}
\author{Wei Cui}
\email{aucuiwei@scut.edu.cn}
\address{School of Automation Science and Engineering, South China University of Technology, Guangzhou 510640, China}
\affiliation{}


\begin{abstract}
	
Quantum computing enables quantum neural networks (QNNs) to have great potentials to surpass artificial neural networks (ANNs). The powerful generalization of neural networks is attributed to nonlinear activation functions. Although various models related to QNNs have been developed, they are facing the challenge of merging the nonlinear, dissipative dynamics of neural computing into the linear, unitary quantum system. In this paper, we establish different quantum circuits to approximate nonlinear functions and then propose a generalizable framework to realize any nonlinear quantum neuron. We present two quantum neuron examples based on the proposed framework. The quantum resources required to construct a single quantum neuron are the polynomial, in function of the input size. Finally, both IBM Quantum Experience results and numerical simulations illustrate the effectiveness of the proposed framework. 

\end{abstract}


\maketitle

\section{\label{Sec1}Introduction}
Artificial neural networks are intelligent computing models vaguely inspired by the biological neural networks that constitute animal brains. Various artificial neural network models have been widely applied in diverse fields such as computer vision, speech recognition, machine translation, and medical diagnosis \cite{nielsen2015neural}. As a new type of computing paradigm with quantum properties, quantum computing has been exhibited to be exponentially or geometrically faster than classical counterparts to solve certain computational problems \cite{365700, PhysRevLett.79.325, PhysRevLett.103.150502}. Combining neural networks with quantum computing, QNNs are considered to be superior to the ANNs in memory capacity, information processing speed, network scale, stability, and reliability\cite{Ezhov2000}. There have been many approaches to build QNNs from different perspectives \cite{schuld2014quest, jeswal2019recent}. In recent years, connecting quantum circuit models to neural network architectures has attracted extensive research interests \cite{schuld2015simulating, da2016quantum, cao2017quantum, wan2017quantum, Wei2018Towards, PhysRevA.100.012334, tacchino2019artificial, cong2019quantum, shao2020quantum, PhysRevA.101.032308}. 

Even the most complicated neural networks are built by a regular connection of identical units called neurons. Usually, a neuron includes two operations: one is the inner product and the other is represented by an activation function. The inner product is a linear operation and generally the activation function is nonlinear. Linear operations are easily realized with quantum computing. As for activation functions, most quantum neurons would like to use the threshold function due to its easy implementation. Ref.~\cite{cao2017quantum} proposes a type of quantum neuron that can simulate a sigmoid-like nonlinear function in light of developed repeat-until-success (RUS) techniques \cite{10.5555/2685179.2685181, 10.5555/3179320.3179329, PhysRevLett.114.080502}. Ref.~\cite{Wei2018Towards} further proposes a non-periodic nonlinear activation function based on RUS circuit and its quantum neuron can be trained with efficient gradient descent. Ref.~\cite{8848854} theoretically proposes an architecture to realize any nonlinear quantum neuron using quantum phase estimation. More existing solutions to nonlinear quantum neurons include the quadratic form of the kinetic term \cite{behrman2000simulations}, dissipative quantum gates \cite{KAK1995143} and reversible circuits \cite{wan2017quantum}, etc. Although there has been a lot of discussions about the nonlinear quantum neurons, these solutions are restricted to exhibit full nonlinear properties except Ref.~\cite{8848854}. However, {Ref.~\cite{8848854} does not give full play to the quantum advantage to obtain well-performed quantum neurons, and it still has the problem of high quantum resource cost.}

To explore universal methods of building any well-functioning and resource-saving quantum neuron, we establish different quantum circuits to approximate nonlinear functions. Moreover, a quantum framework with strong generalization is proposed to obtain any nonlinear quantum neuron. We present two quantum neuron examples based on the proposed framework. The neurons satisfy the criteria proposed by Ref.~\cite{schuld2014quest} in a way that naturally combines the advantages of quantum computing and neuron networks.  The quantum resources required to construct a single quantum neuron are the polynomial functions to the size of the input. Finally, both IBM Quantum Experience results and numerical simulations illustrate the effectiveness of the proposed framework.

The paper is organized as follows. Section~\ref{Sec2} describes two circuits of approximating nonlinear functions. In Section~\ref{Sec3}, we propose a generalized framework to implement any nonlinear quantum neuron and present the analysis of neuron examples. Section~\ref{Sec4} verifies our results by IBMQ experiments and numerical simulations. Conclusions and discussions are presented in Section~\ref{Sec5}.

{\bfseries Notation}. $ X,Y $ and $ Z $ are Pauli matrices. $ [\alpha, \beta] $ indicates an interval from decimal $ \alpha $ to decimal $ \beta $. The norm $ \left\| \cdot \right\| $ always refers to the 2-norm of vectors. The bold italic $ \boldsymbol {i} $ represents the imaginary unit of the complex field. $ x^{T} $ is the transpose of the column vector $ x $. $ y^{D} $ indicates the decimal representation of the binary $ y $. The symbol $ \oplus $ represents the bitwise XOR. $ FT^{\dagger} $ is the inverse quantum Fourier transform. {The $ S $ gate and  $ R_{z}(\alpha) $ gate are defined as $ S = \left[ \begin{matrix} 1 & 0 \\ 0 & \boldsymbol {i} \end{matrix} \right] $, $ R_{z}(\alpha)= \left[
\begin{matrix}
1 & 0 \\
0 & e^{2 \pi \boldsymbol {i} \alpha } 
\end{matrix} \right] $.}

\section{\label{Sec2}Quantum Circuits of Approximating Nonlinear Functions}

The basic task of a computer is assigning values to Boolean functions, which is to give one-bit output for $ n $-bit input \cite{benenti2004principles}. Digital computers compute any complicated function by combining such Boolean functions. Similarly, quantum computers can do the task with black-boxed quantum oracles. There has been some research on quantum oracles \cite{Oracle1994, bennett1997strengths, PhysRevA.65.050304, johansson2019quantum}. An oracle may be a circuit, a physical device, or a pure theory, which helps transform a system from a state to another state.

For a general Boolean function $ f: \left\{0,1\right\}^n \rightarrow \left\{0,1\right\}^m $, {when $ m = 1 $, a standard oracle $ S_{f} $ is defined to act on two input states and return two outputs, 
\begin{eqnarray}
S_{f}:\ket{x}\ket{y} \mapsto \ket{x}\ket{y \oplus f(x)},
\end{eqnarray}
where $ \ket{x} \in (\mathbb{C}^{2})^{\otimes n} $, $ \ket{y} \in \mathbb{C}^{2} $. The following two oracles are examples of standard oracles. The function $ f_{D}: \left\{0,1\right\} \rightarrow \left\{0,1\right\} $ is calculated by Deutsch's oracle
\begin{eqnarray}
S_{f_{D}}:\ket{x}\ket{-} \mapsto \ket{x}\ket{-\oplus f_{D}(x)},
\label{eq:S_{f_{D}}}
\end{eqnarray}
where $ \ket{-\oplus f_{D}(x)} = e^{\pi \boldsymbol {i} f_{D}(x)}\ket{-} $. The function $ f_{G}: \left\{0,1\right\}^{n} \rightarrow \left\{0,1\right\} $ is calculated by Grover's oracle 
\begin{eqnarray}
S_{f_{G}}:\ket{x}\ket{-} \mapsto \ket{x}\ket{-\oplus f_{G}(x)},
\label{eq:S_{f_{G}}}
\end{eqnarray}
where $ \ket{-\oplus f_{G}(x)}= e^{\pi \boldsymbol {i} f_{G}(x)}\ket{-} $ . } 

{When the Boolean function $ f $ satisfies $ m \geq 2 $ , we present the following general oracle 
\begin{eqnarray}
P_{f}: \ket{x}\ket{0\cdots 01} \mapsto e^{\frac{2\pi \boldsymbol {i}}{2^m}f(x) }\ket{x}\ket{{0\cdots 01}},
\label{eq:P}
\end{eqnarray} 
where $ \ket{0 \cdots 01} \in (\mathbb{C}^{2})^{\otimes m} $ and only the last qubit of the $ \ket{0 \cdots 01} $ is in the state $ \ket{1} $ }. The oracles in Eqs.~(\ref{eq:S_{f_{D}}}),  (\ref{eq:S_{f_{G}}}), and (\ref{eq:P}) have the minimal forms, which we denote as minimal phase oracles. Without any auxiliary qubit, a minimal phase oracle $ M_{f} $ is defined to act on one input state and return one output with a specific phase,
\begin{eqnarray}
M_{f}: \ket{x} \mapsto e^{\frac{2\pi \boldsymbol {i}}{2^m} f(x) }\ket{x}.
\end{eqnarray}

Since $ S_{f} $ and $ P_{f} $ are equivalent to $ M_{f} $ in the sense that they can denote the same Boolean function $ f $, $ S_{f} $ and $ P_{f} $ can be converted to $ M_{f} $. In Table \ref{tab1}, the Boolean functions are needed to be computed by corresponding oracles. {The Deutsch's oracle, Grover's oracle and the general oracle in Eq.~\ref{eq:P} can be transformed to} minimal phase oracles, which are explicitly expressed as diagonal unitary matrices. 

\begin{table}[htbp]
	\caption{Transformations of quantum oracles} 
	\centering
	\addtolength{\tabcolsep}{0.8mm} 
	\setlength{\tabcolsep}{0.8mm}{
	\begin{tabular}{ccc}
		\hline \hline
		&  \multicolumn{1}{c}{\begin{tabular}[c]{@{}c@{}} Boolean functions \end{tabular}}& \multicolumn{1}{c}{\begin{tabular}[c]{@{}c@{}} Minimal phase \\oracles \end{tabular}}   \\ \hline
		\midrule
		Deutsch's oracle  & $ \left\{0,1\right\} \rightarrow \left\{0,1\right\}$ & $ \sum_{x=0}^{2^{n}-1}e^{\pi \boldsymbol {i} f_{D}(x)}\ket{x}\bra{x} $  \\ 
		Grover's  oracle & $ \left\{0,1\right\}^n \rightarrow \left\{0,1\right\}$ & $\sum_{x=0}^{2^{n}-1}e^{\pi \boldsymbol {i} f_{G}(x)}\ket{x}\bra{x} $   \\ 
		 \begin{tabular}[c]{@{}c@{}}The oracle in \\Eq.~(\ref{eq:P}) \end{tabular} & $ \left\{0,1\right\}^n \rightarrow \left\{0,1\right\}^m $ & $ \sum_{x=0}^{2^{n}-1}e^{\frac{2\pi \boldsymbol {i}}{2^m} f(x) }\ket{x}\bra{x} $  \\ 
		\hline \hline
	\end{tabular}}
	\label{tab1}
\end{table}

\begin{figure}[h]
	\centering
	\includegraphics[scale=0.48]{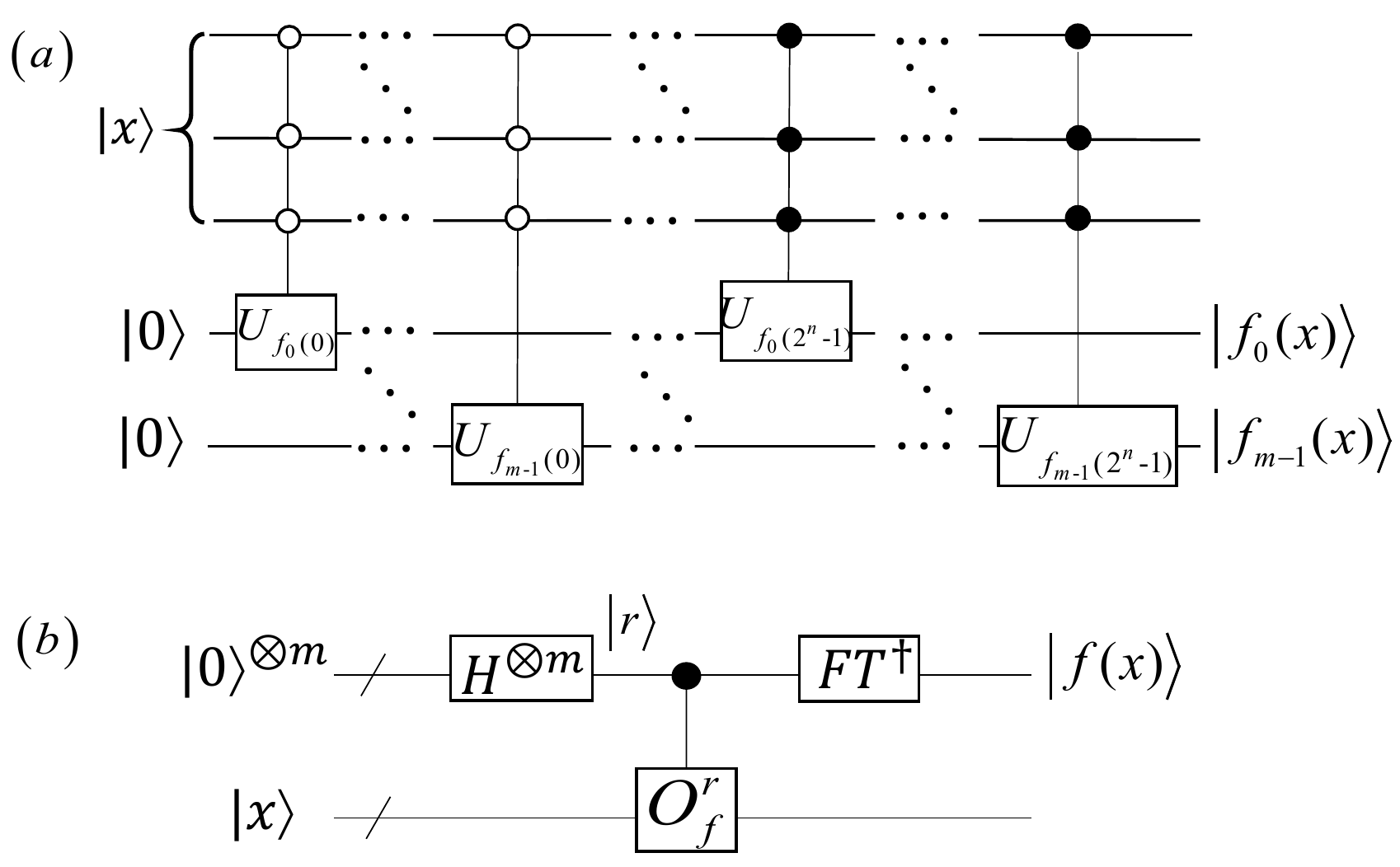}
	\caption{{\bfseries ($ a $)} and {\bfseries ($ b $)} are two quantum circuits for implementing $ U_{f} $ in Eq.~(\ref{eq:U_{f}}). }
	\label{fig:function}
\end{figure}

In Fig.~\ref{fig:function}, we demonstrate two natural quantum circuits to implement the Boolean function $ f: \left\{0,1\right\}^n \rightarrow \left\{0,1\right\}^m $ with a {general} oracle
\begin{eqnarray}
U_{f}:\ket{x}\ket{0}^{\otimes m} \mapsto \ket{x}\ket{f(x)}.
\label{eq:U_{f}}
\end{eqnarray}
After $ U_{f} $ acting on two input states $ \ket{x}\ket{0}^{\otimes m} $,  we could recover
\begin{eqnarray}
f(x)= f_{m-1}(x) \times 2^{m-1} + \cdots + f_{0}(x)\times 2^{0}
\label{eq:f(x)}
\end{eqnarray}
by introducing measurements to the second output state. For $i = 0, 1, \cdots, m-1 $, $ f_{i}(x)  \in \{0, 1\}$ denotes the binary expansion of the $ f(x) $. {Eq.~(\ref{eq:f(x)}) can be regarded as a nonlinear function, whose binary input and first $ m $ bits of binary output are equal to $ x $ and $ f(x) $, respectively. Thus, by Eqs.~(\ref{eq:U_{f}}) and (\ref{eq:f(x)}), we note that Boolean functions approximate corresponding nonlinear functions through quantum oracles.} 

Fig.~\ref{fig:function}\textcolor{blue}{($ a $)} shows the first quntum circuit of $ U_{f} $ in Eq.~(\ref{eq:U_{f}}), where $ U_{f_{i}(j)}\in \{I, X\} $ for $  j = 0, 1,  \cdots, 2^{n}-1 $. The circuit in Fig.~\ref{fig:function}\textcolor{blue}{($ a $)} can be understood as a process of function assignment. Each bit of the input string $ x $ controls each bit of the output string $ f(x) $. Then a specific output $ f(x) $ {is carried out}.

To explore the second quantum circuit of $ U_{f} $ in Eq.~(\ref{eq:U_{f}}), the basic procedures are stated as follow. {Firstly, we construct} a typical minimal phase oracle 
\begin{eqnarray}
O_{f}=\sum_{x=0}^{2^{n}-1}e^{\frac{2\pi \boldsymbol {i}}{2^m} f(x) }\ket{x}\bra{x},
\end{eqnarray}
which adds a phase factor related with the $ f(x) $ to any input $ \ket{x} $. {Secondly, we use the inverse quantum Fourier transform to recover the $ f(x) $ with a certain precision. Fig.~\ref{fig:function}\textcolor{blue}{($ b $)} shows the second quntum circuit of $ U_{f} $ in Eq.~(\ref{eq:U_{f}}).} The controlled-$ O_{f}^{r} $ gates are treated as a composition of  a series of controlled-$ O_{f}^{2^{s}} $ gates by considering the $ s $th qubit in the first register as the control qubit, where $ s = 0,1, \cdots , m-1 $.

\section{\label{Sec3}Nonlinear Quantum neurons}

\begin{figure}[h]
	\centering
	\includegraphics[scale=0.36]{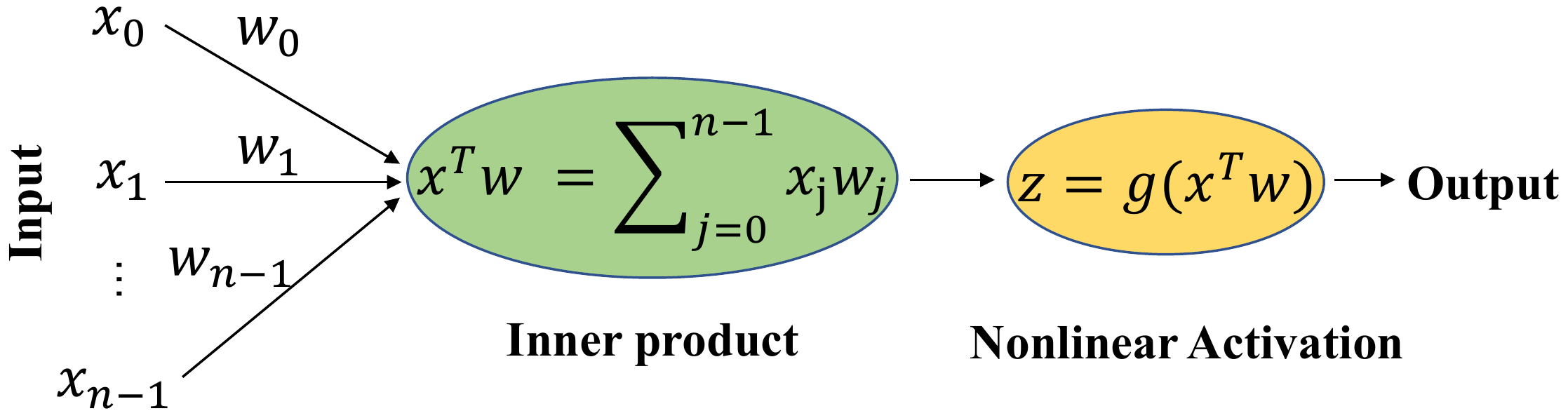}
	\caption{A schematic of a classical neuron.}
	\label{fig:neuron}
\end{figure}

\begin{figure}[htb]
	\centering
	\includegraphics[scale=0.4]{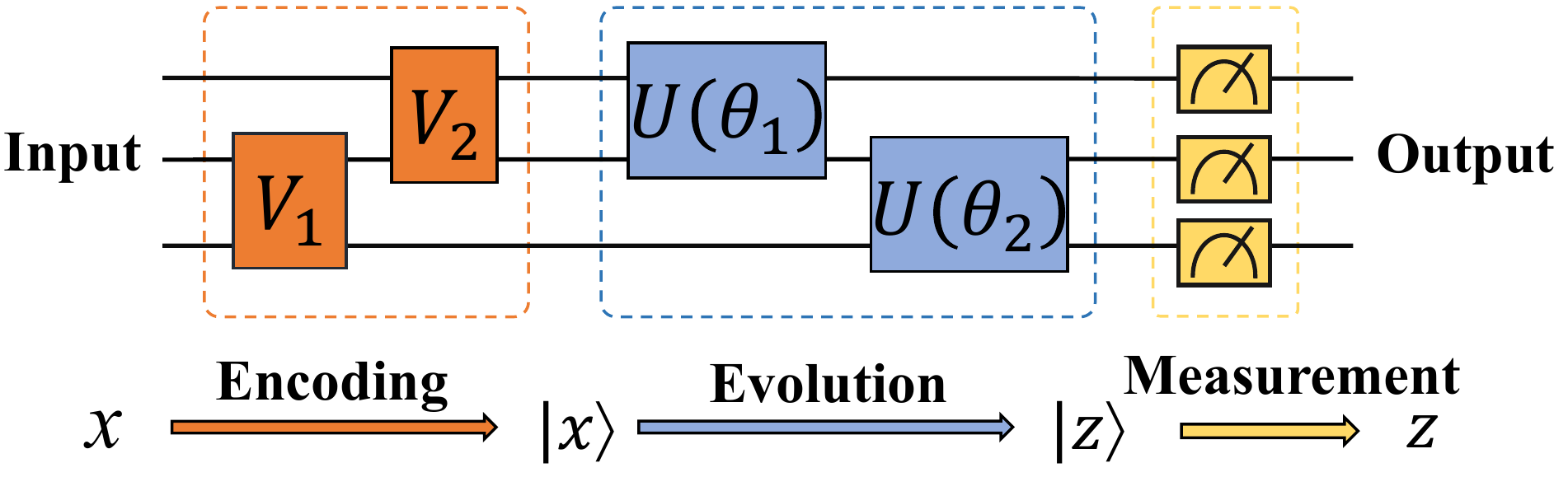}
	\caption{{The proposed framework of implementing} a nonlinear quantum neuron. {$ V_{1} $ represents the initialization of auxiliary qubits. $ V_{2} $ refers to the preparation of quantum dataset. $ U(\theta_{1}) $ {denotes the calculation of the inner product of the input vector and the weight vector.}  $ U(\theta_{2}) $ denotes the approximation of nonlinear functions. $ \theta_{1} $ and  $ \theta_{2} $ are the circuit parameters.}}
	\label{fig:pipeline}
\end{figure}

Deep learning needs a lot of computation resources to train a model. As the amount of transistors in silicon chips approaches the physical limit, QNN is a potential solution to deal with massive and ultra-high dimensional data. However, a recognized definition of QNN has not been proposed on account of the different dynamics between ANNs and quantum computing.

{Fig.~\ref{fig:neuron} shows the schematic of a classical neuron.} As the fundamental building block of ANNs, a neuron classically maps an input vector $ x=(x_{0},\cdots,x_{n-1})^{T} \in \mathbb{R}^{n}$ to an output $ z=g(x^{T}w) $, where $ w=(w_{0},\cdots,w_{n-1})^{T} \in \mathbb{R}^{n}$ is the weight vector. { $ x^{T} w $ is the inner product} and $ g $ is usually a nonlinear activation function. 

Motivated by this schematic, we propose a generalizable framework {of implementing} nonlinear quantum neurons in Fig.~\ref{fig:pipeline}. {The framework maps the existing neuron designed for classical computers to quantum circuits and} achieves the nonlinear mapping of classical data through three processes: encoding, evolution, and measurement. The encoding is to represent the features of classical data by a quantum system. Following the postulates of quantum mechanics, the evolution maps a quantum state $ \ket{x} $ onto $ \ket{z} $ with unitary operators, {which are the key to successfully implement nonlinear quantum neurons.}  Reducing the dimension of the state space, the measurement is necessary to induce nonlinear quantum neurons. When quantum neurons are connected to build QNNs, the measurement of each neuron in all middle layers should be postponed until the last layer. 

\subsection{\label{Sec3A}Neuron Examples}

Let $ \mathcal{M}~\triangleq~ \{x^{i}:i = 1,2, \cdots,q \} \in \mathbb{R}^{n} $ be a classical dataset. We denote a sample vector as $ x^{i}=(x_{0}^{i},x_{1}^{i},\cdots,x_{n-1}^{i})^{T} $, where $ x_{j}^{i} $ for $ j = 0, 1, \cdots, n-1 $ represents the $ j $th component value of the $ i $th sample in $ \mathcal{M} $. The $ \mathcal{M} $ has been processed to ensure that each $ x_{j}^{i} $ satisfies $ 0 \le x^{i}_{j} <1 $. Ref.~\cite{schuld2018supervised} has systematically introduced encoding methods and discussed state preparation routines. Since data can be encoded into {computational basis states} or amplitudes, we apply basis encoding and amplitude encoding to our framework.

\begin{figure}[htb]
	\centering
	\includegraphics[scale=0.43]{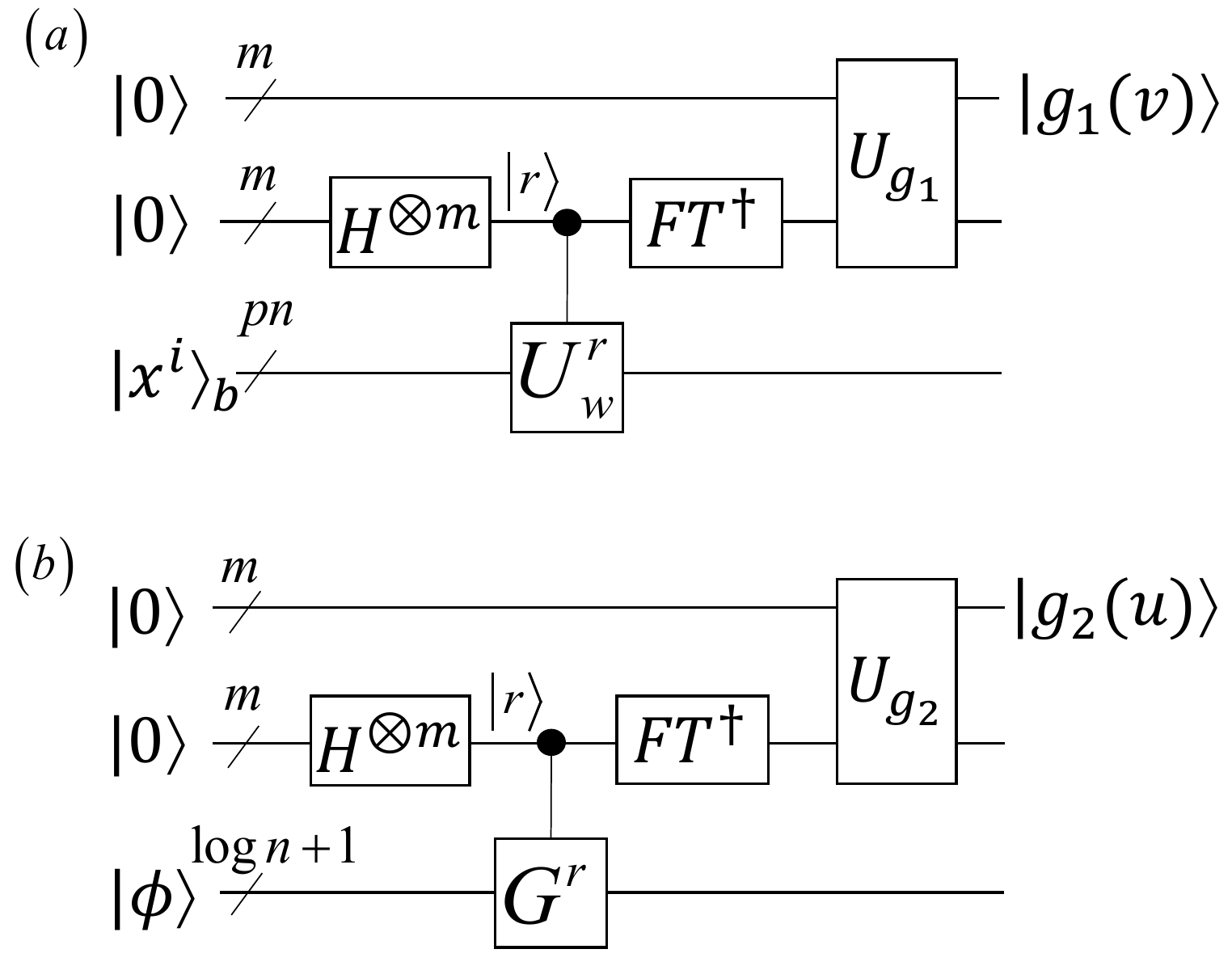}
	\caption{{\bfseries ($ a $)} and {\bfseries ($ b $)} are represented nonlinear quantum neuron examples based on basis encoding and amplitude encoding, respectively.}
	\label{fig:example}
\end{figure}

Fig.~\ref{fig:example} shows two nonlinear quantum neuron examples. Both neurons are composed of three quantum registers, which are labeled as the first register, the second register and {the} third register {(from top to bottom).} If we perform measurements on the first register, we can obtain {the output of a quantum neuron.} The second register is an auxiliary register to calculate the inner product  {of the input vector and the weight vector.} The third register is the sample encoding register, which is used to encode the classical data into quantum states. The parameter $ m $ is an integer that always relates to precision.  {In the following analysis,} we denote the weight vector $ w $ to compute inner product with the sample $ x^{i} $.  

Fig.~\ref{fig:example}\textcolor{blue}{(a)} is an example of nonlinear quantum neurons based on basis encoding. Basis encoding transforms each $ x^{i} $ into a product state 
\begin{eqnarray}
\ket{x^{i}} _{b} = \ket{x_{0}^{i},x_{1}^{i}, \cdots, x_{n-1}^{i}} \in (\mathbb{C}^{2})^{\otimes pn},
\end{eqnarray}
{where $ p $ is a fixed number of qubits for a given precision to approximate the component} $ x_{j}^{i} $, $ x_{j}^{i}  = x_{j_{1}}^{i} / 2^{1} + \cdots + x_{j_{p}}^{i} / 2^{p}$, $ x_{j_{k}}^{i} \in \{0, 1\}$ for $ k = 1, 2, \cdots, p $. By encoding classical data into an orthonormal basis of the Hilbert space, quantum parallelism allows a linear combination of the basis to represent $ \mathcal{M} $ in only one quantum state, {which means} it is possible to process all data in parallel. 

When the controlled-phase gate $ CR_{z}(\frac{\alpha}{2^{m}}) $ acts on two quantum state $ \ket{+}\ket{x_{j_{k}}^{i}} $, we obtain
\begin{equation}
CR_{z}(\frac{\alpha}{2^{m}})(\ket{+}\ket{x_{j_{k}}^{i}}) =\dfrac{1}{\sqrt{2}}(\ket{0}\ket{x_{j_{k}}^{i}}+e^{\frac{2\pi \boldsymbol {i}}{2^m} x_{j_{k}}^{i} \alpha}\ket{1}\ket{x_{j_{k}}^{i}}),
\end{equation}
 {where  $ \frac{\alpha}{2^{m}} $ is a quantum phase factor and $ CR_{z}(\frac{\alpha}{2^{m}}) $ can be written as} 
\begin{equation}
CR_{z}(\dfrac{\alpha}{2^{m}}) = \left[
\begin{matrix}
1 & 0 & 0 & 0\\
0 & 1 & 0 & 0\\
0 & 0 & 1 & 0\\
0 & 0 & 0 & e^{\frac{2\pi \boldsymbol {i}}{2^m} \alpha} 
\end{matrix} \right].
\end{equation}
Thus, {in Fig.~\ref{fig:example}(a)} the controlled-$ U_{w}^{r} $ gates are constructed with a series of $ CR_{z}(2^{-(k+m)}w_{j}) $ gates, which are performed on the $ k $th qubit of the $ j $th component of $ x^{i} $. Meanwhile, the controlled-$ U_{w}^{r} $ gates can be regarded as a composition of a series of controlled-$ U_{w}^{2^{s}} $ gates by considering the $ s $th qubit in the second register as the control qubit. After the application of the controlled-$ U_{w}^{2^{s}} $ on the second and third registers, these two registers become 
$$\dfrac{1}{\sqrt{2^{m}}} \sum_{t=0}^{2^{m}-1}e^{ \frac{2\pi \boldsymbol {i}}{2^{m}} t (x^{i})^{T}w} \ket{t}\ket{x^{i}}_{b}.$$

To apply the inverse quantum Fourier transform  {to} the second register,  {the output of the second register is $ \ket{v} $, where $  v \in \left\{0,1\right\}^m $. Since $ (x^{i})^{T}w $ can be positive or negative, we denote the $ v $ as the original code of a number, whose complement is an approximation of $ (x^{i})^{T}w $.  Then} we use the Boolean function $ g_{1}(v):\left\{0,1\right\}^m \rightarrow \left\{0,1\right\}^m $ to approximate any activation function. The $ g_{1}(v) $ is calculated by the oracle 
\begin{eqnarray}
U_{g_{1}}: \ket{v}\ket{0}^{\otimes m} \mapsto \ket{v}\ket{g_{1}(v)},
\label{eq:v}
\end{eqnarray}
which can be implemented by the circuit in Fig.~\ref{fig:function}\textcolor{blue}{($ a $)} or Fig.~\ref{fig:function}\textcolor{blue}{($ b $)}.

Fig.~\ref{fig:example}\textcolor{blue}{(b)} is an example of nonlinear quantum neurons based on amplitude encoding. Amplitude encoding encodes the sample $ x^{i} $ into an entangled state
\begin{eqnarray}
\ket{x^{i}} _{a} = \frac{1}{\left\| x^{i} \right\|}\sum_{j=0}^{n-1}x_{j}^{i}\ket{j} \in \mathbb{C}^{n}. 
\end{eqnarray}
This {process} only uses log$ n $ qubits to represent a sample consisting of  $ n $ features. 

In Fig.~\ref{fig:example}\textcolor{blue}{(b)}, the weight vector $ w $ is encoded in
\begin{eqnarray}
\ket{w} = \frac{1}{\left\| w \right\|}\sum_{j=0}^{n-1}w_{j}\ket{j} \in \mathbb{C}^{n}.
\end{eqnarray}
{The quantum states $ \ket{x^{i}} _{a} $ and $ \ket{w} $ could be carried out by quantum random access memory~(qRAM) \cite{PhysRevLett.100.160501}, which is an efficient algorithm to prepare quantum state under certain data structure.} The inner product $ \langle w \ket{x^{i}}_{a} $ could be estimated with swap test which has been widely  {used in} quantum machine learning \cite{lloyd2014quantum, zhang2016quantum, shao2020quantum, PhysRevA.100.012334}. Firstly,  {we can} prepare the quantum state 
\begin{eqnarray}
\ket{\phi} = \frac{1}{\sqrt{2}}(\ket{+}\ket{x^{i}}_{a}+\ket{-}\ket{w}),
\end{eqnarray}
which can be rewritten as
\begin{eqnarray}
\ket{\phi} = \frac{1}{2}(\ket{0}(\ket{x^{i}}_{a}+\ket{w})+\ket{1}(\ket{x^{i}}_{a}-\ket{w})).
\end{eqnarray}
The amplitude of $ \ket{0} $ is 
\begin{eqnarray}
\sin \gamma = \left. \sqrt{1+\langle w \ket{x^{i}}_{a}} \middle/ \sqrt{2} \right. 
\end{eqnarray}
and the amplitude of $ \ket{1} $ is
\begin{eqnarray}
\cos \gamma =\left. \sqrt{1- \langle w \ket{x^{i}}_{a}} \middle/ \sqrt{2} \right.,
\label{eq:cos}
\end{eqnarray}
where $ \gamma \in [0,\frac{\pi}{2}] $. With Schmidt decomposition, $ \ket{\phi} $ is decomposed into
\begin{eqnarray}
\ket{\phi} = \dfrac{-\boldsymbol {i}}{\sqrt{2}}(e^{\boldsymbol {i} \gamma}\ket{w_{+}}-e^{-\boldsymbol {i} \gamma}\ket{w_{-}}),
\end{eqnarray}
where 
\begin{eqnarray}
\ket{w_{\pm}}=\dfrac{1}{\sqrt{2}}(\ket{0}(\ket{x^{i}}_{a}+\ket{w})\pm \boldsymbol {i}\ket{1}(\ket{x^{i}}_{a}-\ket{w}) ).
\end{eqnarray}
Secondly,  {we can} construct a unitary transformation 
\begin{eqnarray}
G = (I^{\otimes \log n+1}-2\ket{\phi}\bra{\phi})(Z\otimes I^{\otimes \log n}).
\end{eqnarray}
The eigenvalues of $ G $ are $ e^{\pm \boldsymbol {i} 2 \gamma}  $ and the corresponding eigenvectors are $ \ket{w_{\pm}} $. Then we can use quantum phase eatimation to estimate $ \gamma $. The controlled-$ G^{r} $ gates denote a composition of a series of controlled-$ G^{2^{s}} $ gates by considering the $ s $th qubit in the second register as the control qubit. After applying quantum phase estimation, the output of the second and third register is  { 
\begin{eqnarray}
\ket{\psi} = \dfrac{-\boldsymbol {i}}{\sqrt{2}}(e^{\boldsymbol {i} \gamma} \ket{u}\ket{w_{+}}-e^{-\boldsymbol {i} \gamma} \ket{2^{m}-u^{D}}\ket{w_{-}}),
\end{eqnarray}
where the first qubit of $ \ket{u} $ is in state $ \ket{0} $ and $ u \in \left\{0,1\right\}^m $. $ u^{D}\pi/2^{m-1} $ is an approximation of $ 2\gamma $. By Eq.~(\ref{eq:cos}), we obtain
\begin{equation}
u^{D} \approx \arccos(-\langle w \ket{x^{i}}_{a})2^{m-1}/\pi. 
\end{equation}}
Then we use the Boolean function $ g_{2}(u):\left\{0,1\right\}^m \rightarrow \left\{0,1\right\}^m $ to approximate any activation function. The $ g_{2}(u) $ is calculated by the oracle
\begin{eqnarray}
U_{g_{2}}: \ket{u}\ket{0}^{\otimes m} \mapsto \ket{u}\ket{g_{2}(u)},
\label{eq:u}
\end{eqnarray}
which can be implemented by the circuit in Fig.~\ref{fig:function}\textcolor{blue}{($ a $)} or Fig.~\ref{fig:function}\textcolor{blue}{($ b $)}.

By Eqs.~(\ref{eq:v}) and (\ref{eq:u}), we note that the outputs of the second registers can be seen as the inputs of the  {Boolean functions.} Other existing quantum computing models can apply to our framework. So long as the outputs of the second registers are the bijective functions with respect to $ (x^{i})^{T} w $, a specific oracle would be available to approximate any corresponding activation function.

\subsection{Comparisons}

The generalizable framework has been proposed to implement nonlinear quantum neurons, which are suitable to deal with massive and ultra-high dimensional data. The numbers of qubits and elementary quantum gates measure the quantum resources to implement a quantum neuron. Table \ref{tab2} shows the comparisons of different nonlinear quantum neurons with $ n $ representing the input size (the dimensions of input vectors).

Considering the space cost, the $ m $ and $ p $ are parameters that influence the binary representation precisions of the numbers. For example, the long int types and double-precision floating-point types are represented by fixed 64 bits in a common digital computer. Ref.~\cite{8848854} encodes input vectors and the corresponding weight vectors into the basis states. If each component of the vectors is represented by $ p $ qubits, Ref.~\cite{8848854} needs $ 2pn+m $ qubits to implement its quantum neuron. From the analysis in Section~\ref{Sec3A},  Fig.~\ref{fig:example}\textcolor{blue}{($ a $)} requires $ pn+2m $ qubits for a neuron and Fig.~\ref{fig:example}\textcolor{blue}{($ b $)} needs log$ n+2m+1 $ qubits. When $ n $ is much larger than the parameters $ m $ and $ p $, Fig.~\ref{fig:example}\textcolor{blue}{($ a $)} uses about half fewer qubits than Ref.~\cite{8848854}. Moreover, Fig.~\ref{fig:example}\textcolor{blue}{($ b $)} exponentially reduces the requirement of qubits.

Turning to the time cost, the elementary gates are physically implementable  gates, include single-qubit rotations and entangling two-qubit gates. A general unitary operator can be decomposed into elementary gates following the general principles presented by {Refs}.~\cite{barenco1995elementary, PhysRevLett.93.130502}. We use $ C^{d-1}R_{z} $ gates to represent multiple controlled-$ R_{z} $ gates acting on $ d-1 $ control qubits and one target qubit. A $ C^{d-1}R_{z} $ gate can be simulated by $ O(d^{2}) $ elementary gates or else $ O(d) $ gates with one auxiliary qubit [\cite{barenco1995elementary}, Corollary 7.6]. The accurate evaluation of diagonal unitary operators is often the most resource-intensive element of quantum algorithms. There are some discussion of decomposition methods for arbitrary diagonal unitaries \cite{DBLP:journals/qic/BullockM04, Welch_2014, houshmand2014decomposition}. Generally, an arbitrary $ d $-qubit diagonal unitary operator could be decomposed into $ 2^{d}~C^{d-1}R_{z} $ gates.  Therefore, constructing quantum circuits for diagonal computations generally requires $ O(d^{2}2^{d}) $ elementary gates without any auxiliary qubit. Specially, Ref.~\cite{DBLP:journals/qic/BullockM04} provides circuits of size $ O(2^{d}) $ for arbitrary $ d $-qubit diagonal unitaries, which is the best-known compiling algorithm. An efficient implementation of the phase estimation includes $ O(m^{2}) $  elementary gates for an inverse quantum Fourier transform and one call to controlled unitary operator black box \cite{nielsen2010quantum}. Here, we would discuss the elementary gate complexity of constructing a black box rather than the query complexity. Ref.~\cite{8848854} consists of arbitrary $ 2pn $-qubit diagonal operations, which should be decomposed into $ O(2^{2pn}) $ elementary gates with the asymptotically optimal circuits of standard techniques \cite{DBLP:journals/qic/BullockM04}. Therefore, Ref.~\cite{8848854} needs $ O(m2^{2pn}+m^{2}) $ gates to implement a single neuron. In our cases, the corresponding unitary operator black box in Fig.~\ref{fig:example}\textcolor{blue}{($ a $)} and Fig.~\ref{fig:example}\textcolor{blue}{($ b $)} are the $ U_{w}^{r} $ gate and the $ G^{r} $ gate, respectively. The analysis in Section~\ref{Sec3A} shows that the $ U_{w}^{r} $ gate can exactly be decomposed into $ pn~R_{z} $ gates. The $ G^{r} $ gate can be expressed as $ O(n^{2}) $ elementary gates according to the decomposition method for general multi-qubit gates in Ref.~\cite{PhysRevLett.93.130502}. If $ U_{g_{1}} $ and $ U_{g_{2}} $ are implemented by Fig.~\ref{fig:function}\textcolor{blue}{($ b $)}, the gate complexity is $ O(m2^{m}) $. In summary, the total gate complexity of the Fig.~\ref{fig:example}\textcolor{blue}{($ a $)} and Fig.~\ref{fig:example}\textcolor{blue}{($ b $)} is $ O(pn+m2^{m}) $ and  $ O(mn^{2}+m2^{m}) $, respectively.

As mentioned before, the parameters $ m $ and $ p $ represent the precisions of the inputs and outputs of neurons, respectively. In neural networks, the activation output of a neuron can be regarded as the input of the neuron connected to it. Therefore, the values of $ m $ and $ p $ can be considered equal (or approximately equal). In this way, our quantum neurons always use fewer quantum resources than Ref.~\cite{8848854}, even exponentially reduce the costs when $ n $ is much larger than $ m $ and $ p $.

\begin{table}[htbp]
	\caption{Comparison of different quantum neurons} 
	\centering
	\addtolength{\tabcolsep}{1mm} 
	
	\setlength{\tabcolsep}{1.3mm}{
	\begin{tabular}{lccc}
		\hline \hline
		\begin{tabular}[c]{@{}c@{}}Quantum \\  neurons\end{tabular} & \begin{tabular}[c]{@{}c@{}}Input\\ Features\end{tabular} & \begin{tabular}[c]{@{}c@{}}Number\\ of qubits\end{tabular} & \begin{tabular}[c]{@{}c@{}} Gate \\ complexity \end{tabular} \\ \hline
		\midrule
		Fig.~\ref{fig:example}\textcolor{blue}{($ a $)}      &  Binary          &    $ pn+2m $         &     $ O(pn+m2^{m}) $         \\
		Fig.~\ref{fig:example}\textcolor{blue}{($ b $)}      &   Continuous     & $  $log$ n +2m+1 $     &      $ O(mn^{2}+m2^{m}) $        \\
		Ref.~\cite{8848854}          &   Binary           & $ 2pn+m $        & $ O(m2^{2pn}+m^{2}) $                     \\                                           
		\hline \hline
	\end{tabular}}
	\label{tab2}
\end{table}

\section{\label{Sec4}Experiments}

\subsection{IBMQ Experiments}

\begin{figure*}[htb]
	\centering
	\includegraphics[scale=0.46]{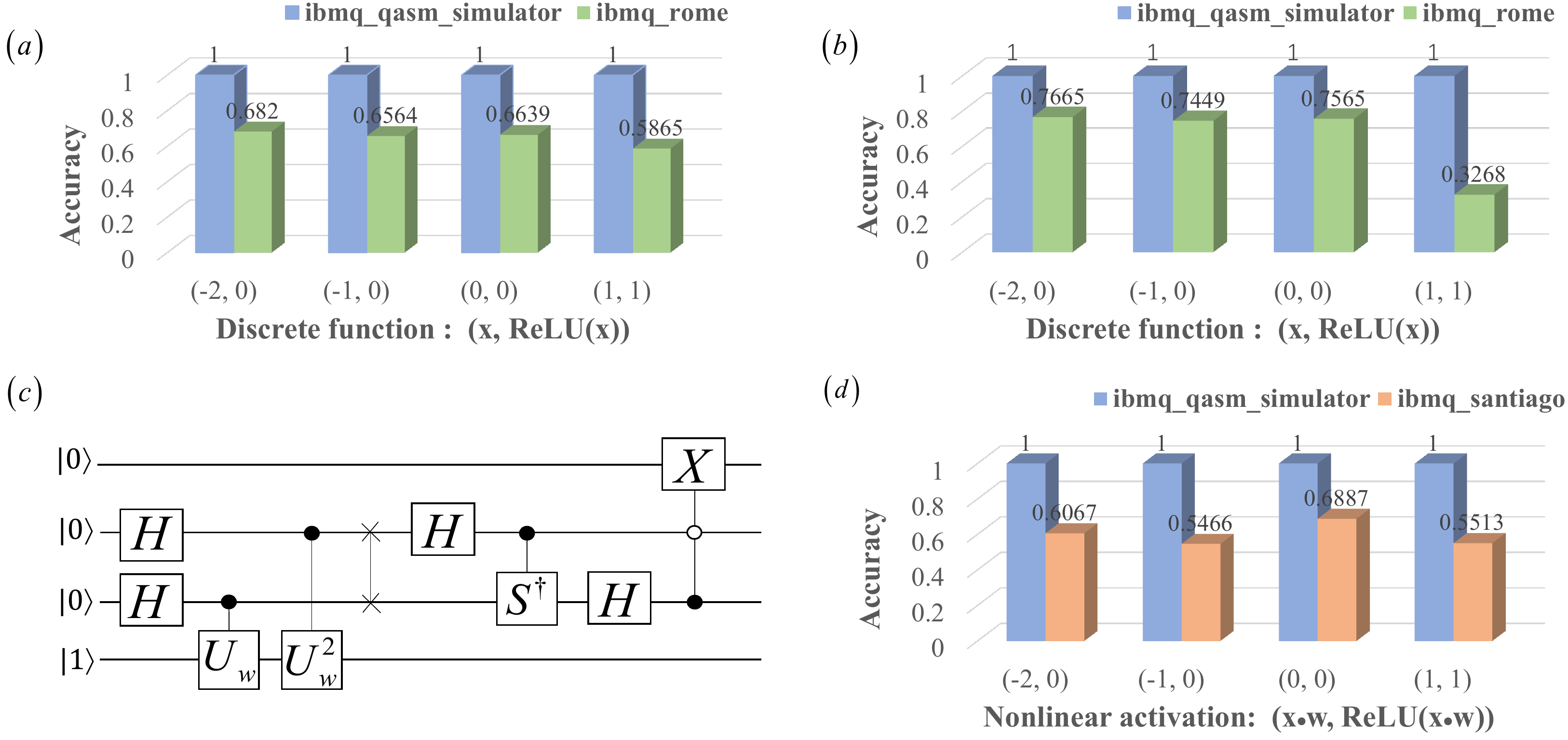}
	\caption{Implementing discrete ReLU with different quantum circuits. {All circuits are executed 8192 times {(the maximum allowed)} on IBM’s quantum simulator and quantum computers. The accuracy indicates the proportion of the correct outputs of all outputs for given inputs.  {\bfseries ($ a $)} The results of using 4 qubits to implement discrete ReLU with the circuit in Fig.~\ref{fig:function}\textcolor{blue}{($ a $)}.  {\bfseries ($ b $)} The results of using 4 qubits to implement discrete ReLU with the circuit in Fig.~\ref{fig:function}\textcolor{blue}{($ b $)}. {\bfseries ($ c $)} An example of a typical quantum circuit for a  nonlinear quantum neuron based on the Fig.~\ref{fig:example}\textcolor{blue}{($ a $)} with 4 qubits, where the sign bit of the activation output is ignored due to ReLU's nonnegative output. In this example, the input data is $ x = 1 $ and the $ U_{w} = R_{z}(w/4) , w \in \{-2,-1,0,1\} $. {\bfseries ($ d $)} The running results of the circuit in Fig.~\ref{fig:IBM}\textcolor{blue}{($ c $)}.}}
	
	\label{fig:IBM}
\end{figure*}

Nowadays, Noisy Intermediate-Scale Quantum (NISQ) technology has been available  \cite{Preskill2018quantumcomputingin}. IBM quantum experience (IBMQ for short), a cloud-based quantum computer, has become a {popular platform for quantum computing.} We implement the discrete Rectified Linear Unit (ReLU) activation function with different quantum circuits based on the platform.

Fig.~\ref{fig:IBM}\textcolor{blue}{($ a $)} and Fig.~\ref{fig:IBM}\textcolor{blue}{($ b $)} shows the results of running the two circuits in Fig.~\ref{fig:function}. Both the circuits are implemented with 4 qubits, where half the qubits are for inputs and the other half are for outputs. The first qubit of inputs (and outputs) is regarded as the sign of a number and the second represents the one-bit integer. Then the domain and codomain of discrete ReLU can be computed classically. The input states of the two circuits are initialized to superposition states with equal weights. {Fig.~\ref{fig:IBM}\textcolor{blue}{($ c $)} presents an example of a nonlinear quantum neuron based on the Fig.~\ref{fig:example}\textcolor{blue}{($ a $)}, and its running results are shown in Fig.~\ref{fig:IBM}\textcolor{blue}{($ d $)}.} 

{Since the inputs and outputs of our typical circuits are integers, they can be written in a binary form precisely, and the errors are only caused by the quantum gate errors and readout errors.} We obtain 100\% accuracies on {\bfseries ibmq\_qasm\_simulator} backend. {As the gate errors and readout errors on real quantum hardware reduce the executed reliability of quantum circuits, the accuracies on {\bfseries ibmq\_rome} backend and {\bfseries ibmq\_santiago} backend are about 70\% and 60\%, respectively.} 

From the experimental results, we conclude that the two circuits in Fig.~\ref{fig:function} have similar performance in the case of using the same number of qubits. The results on the simulator verify the feasibility of the circuits, specifically including the circuit of a typical quantum neuron with the ReLU activation function. The results on quantum computers illustrate that our specific circuits still have a positive performance on real quantum hardware in the NISQ era of quantum computing. Certainly, the accuracies on quantum computers can be improved with the circuit optimization theory and well performance of the real quantum hardware. We just show the mapping experiment of discrete ReLU. It can be extended to any type of function.

\begin{figure*}[htb]
	\centering
	\includegraphics[scale=0.52]{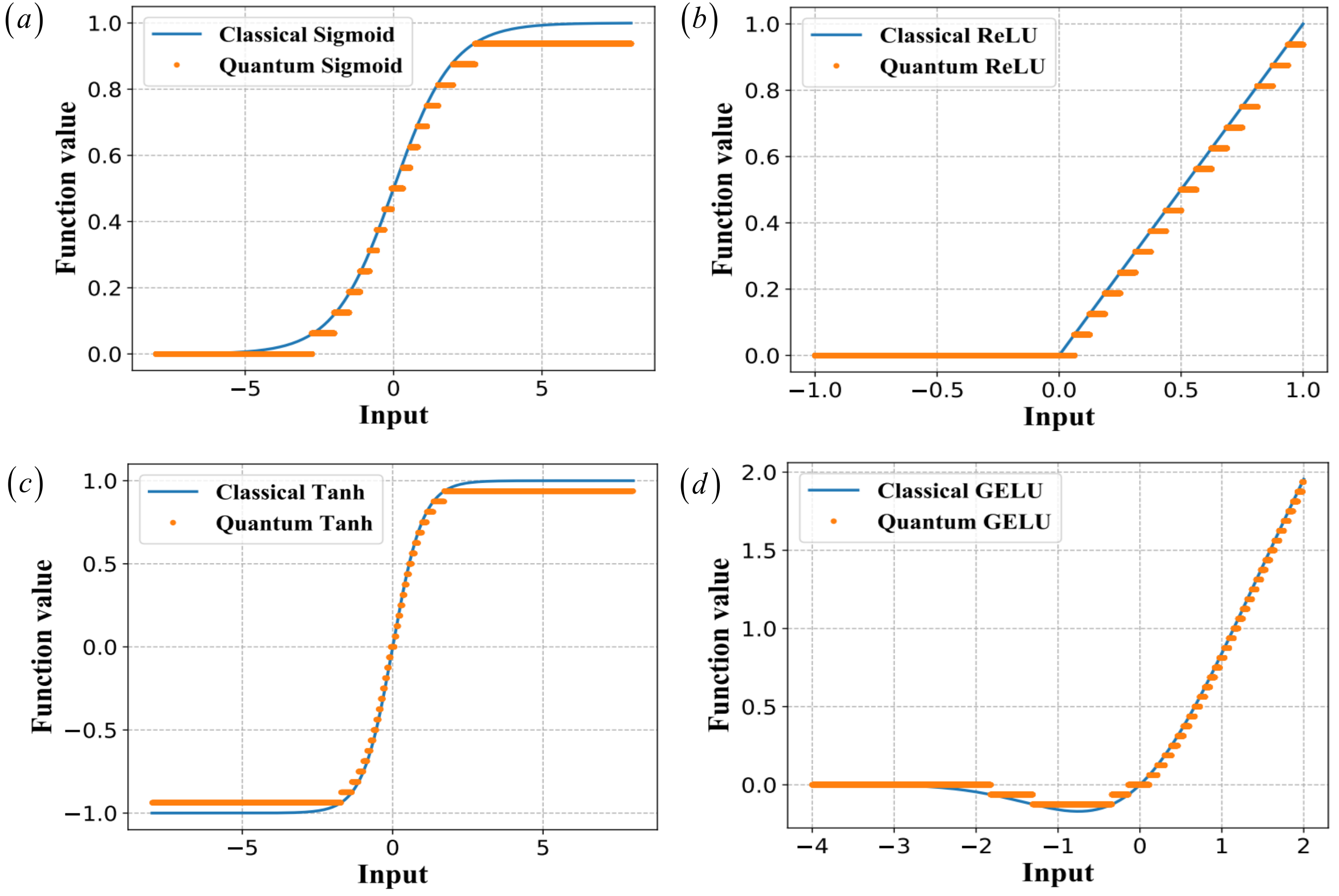}
	\caption{Different types of activation functions: {\bfseries ($ a $)} Sigmoid, {\bfseries ($ b $)} ReLU, {\bfseries ($ c $)} Tanh, and {\bfseries ($ d $)} GELU.}
	\label{fig:numerical}
\end{figure*}

\subsection{Numerical Simulations of A Quantum Neuron}
The performance of ANNs is significantly related to the selection of activation functions. In many scenarios, it is necessary to trade off both the advantages and disadvantages of different activation functions. {There are some} commonly used classical activation functions, such as Sigmoid, Hyperbolic Tangent (Tanh), Gaussian Error Linear Units (GELU), and so on. Fig.~\ref{fig:numerical} shows the results of simulating these activation functions based on the nonlinear quantum neuron proposed in Fig.~\ref{fig:example} {($ a $)}, {where the input is the value of $ (x^{i})^{T} w $.} The neuron has two approximate calculations: one is the approximation of $ (x^{i})^{T}w $ and the other is the approximation of the activation function, which is computed by the classical 64-bit computer. These two approximations are carried out by the first and second registers, respectively. We stipulated that both of these two approximations accurate to eight qubits, one qubit for representing the sign, three qubits for integers, and four qubits for decimals.

The results exhibit that the activation functions computed by the nonlinear quantum neuron can simulate classical activation functions with a few qubits. In the numerical simulations, we only consider the ideal case where the quantum phase estimation runs successfully and returns accurate results. So the approximate error is only caused by the insufficient number of output qubits. Actually, the errors would infinitely approach zero as the number of qubits increases. 

\subsection{Numerical Simulation of A Simple QNN}

\begin{figure*}[htb]
	\centering
	\includegraphics[scale=0.5]{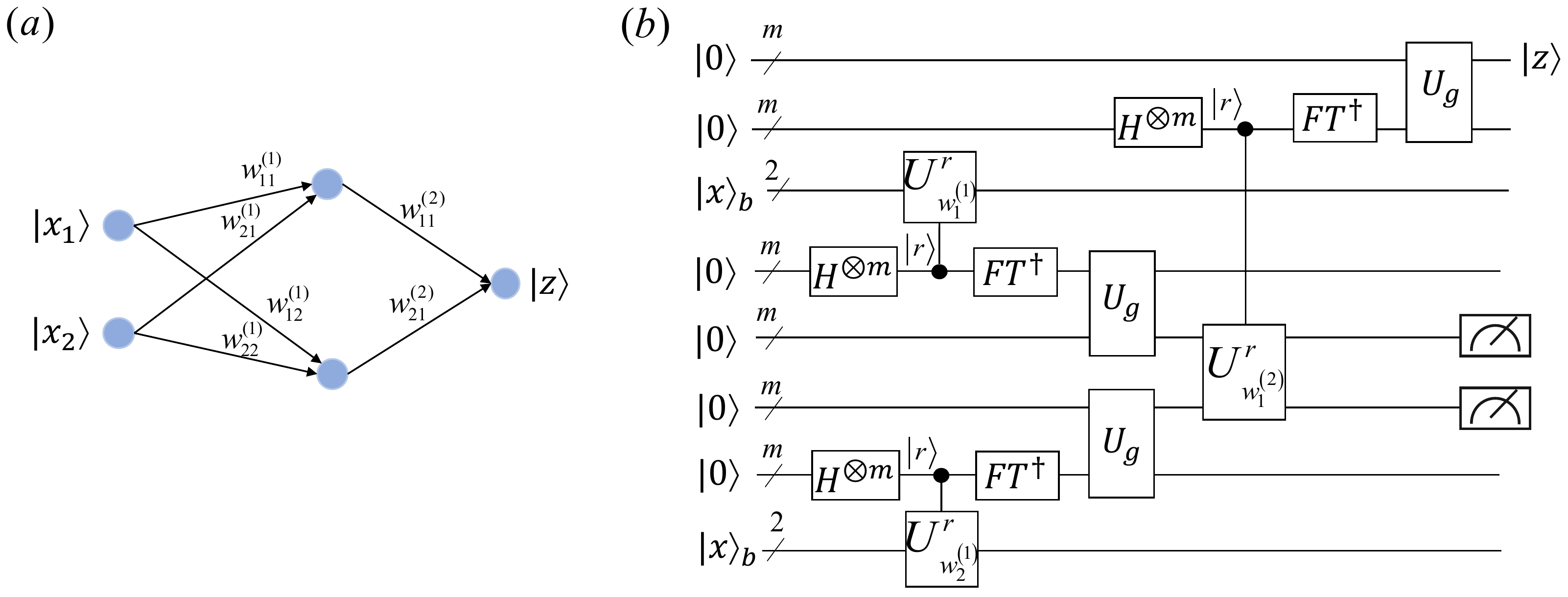}
	\caption{ Construction of a simple quantum feedforward neural network with the proposed nonlinear quantum neurons. Here the input states are  $ \ket{x_{1}} $ and $ \ket{x_{2}} $, and the output state is $ \ket{z} $. {\bfseries ($ a $)} Quantum feedforward neural network model. {\bfseries ($ b $)} The quantum feedforward neural network represented by a circuit, where $  \ket{x}_{b} = \ket{x_{1}} \otimes \ket{x_{2}} $, $ U_{w^{(1)}_{i}} = R_{z}(w^{(1)}_{i1}) \otimes R_{z}(w^{(1)}_{i2}) $, $ U_{w^{(2)}_{i}} = R_{z}(w^{(2)}_{i1}) \otimes R_{z}(w^{(2)}_{i2})\otimes \dots \otimes R_{z}(w^{(2)}_{im}) $. }
	\label{fig:qnn}
\end{figure*}

\begin{figure}[htb]
	\centering
	\includegraphics[scale=0.53]{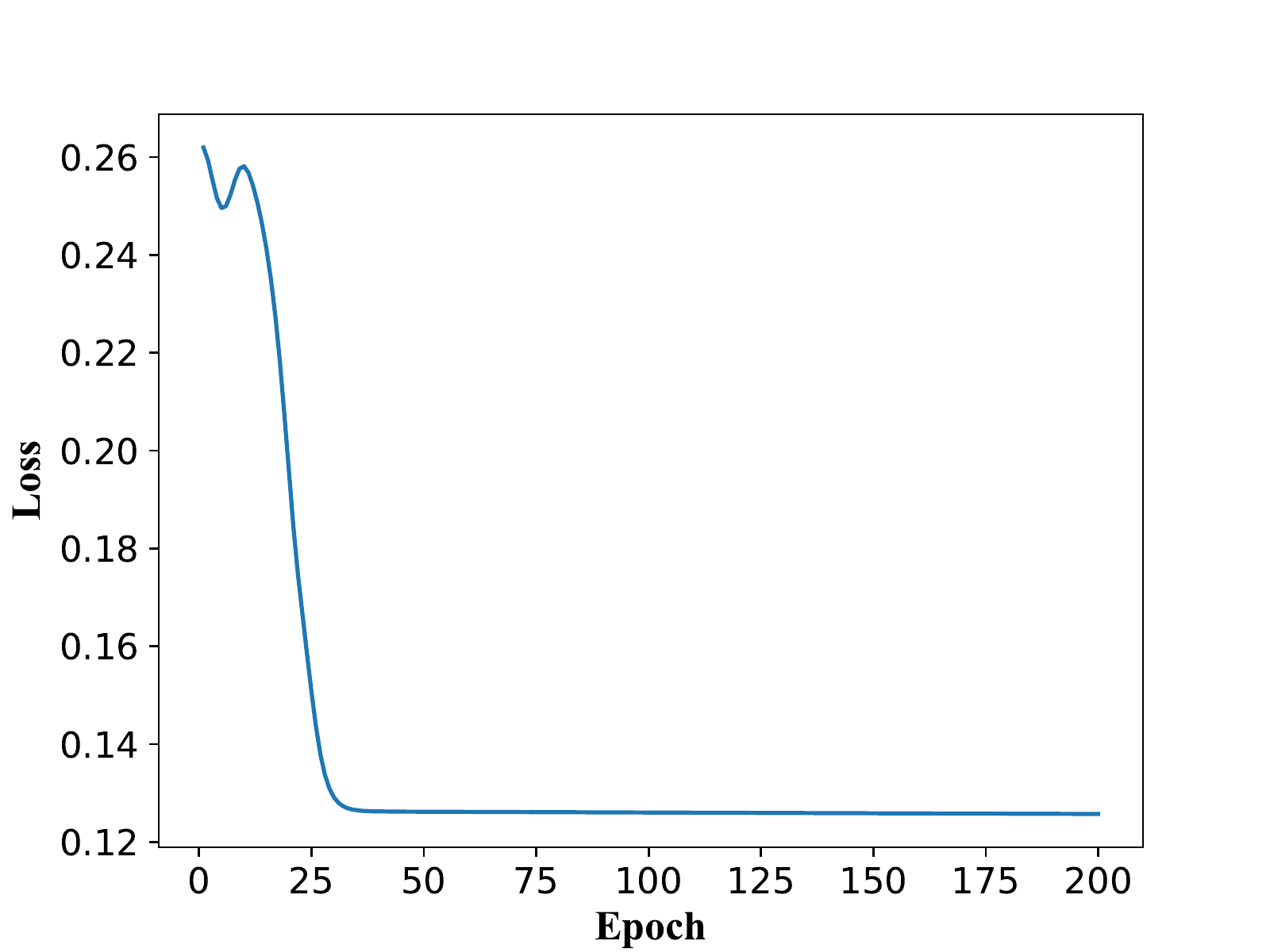}
	\caption{The learning curve.}
	\label{fig:learningcurve}
\end{figure}

As the proposed quantum neurons are proved to be able to simulate any nonlinear activation function, one of their common usages is similar to classical neurons. The quantum neurons of one layer are connecting to quantum neurons of the immediately preceding and immediately following layers. Fig.~\ref{fig:qnn} shows the construction of a quantum feedforward neural network (QFNN) with three quantum neurons proposed in Fig.~\ref{fig:example}\textcolor{blue}{($ a $)}, where two neurons are for the hidden layer and the other one for the output layer. The QFNN has two inputs, which constitute the input layer, i.e., the zeroth layer, of the neural network. For $ k=1,2 $, the $ j $th neuron in the $ k $th layer and the $ i $th neuron in the $ (k-1) $th layer are connected by an edge $ w_{ij}^{(k)} $, which belongs to the set $ \{w_{11}^{(1)}, w_{12}^{(1)}, w_{21}^{(1)}, w_{22}^{(1)}, w_{11}^{(2)}, w_{12}^{(2)}\} $.

With the quantum Sigmoid activation function implemented by the oracle $ U_{g} $, the three-neuron QFNN was trained with backpropagation to solve the well-known XOR problem, which consists of four samples ($ q $ = 4). During the training process, we applied the batch gradient descent algorithm to minimize the following mean square loss function
\begin{eqnarray}
\mathcal{L}=\dfrac{1}{q}\sum_{i=1}^{q} (d^{i}-z^{i})^{2},
\end{eqnarray}
where for the $ i $th sample, $ d^{i} $ is the desired output and $ z^{i} $ is the result of performing measurement on the output $ \ket{z^{i}} $ of the QFNN. The update rule for the weights is
\begin{eqnarray}
w_{(l+1)}=w_{(l)} - \eta\dfrac{\partial \mathcal{L}}{w_{(l)}},
\end{eqnarray}
where $ w_{(l+1)} $ is the weight after the $ (l + 1) $th step of the iteration process and $ \eta $ is an adjustable positive step length. The learning curve is presented in Fig.~\ref{fig:learningcurve}, from which we can see that the loss converges to about 0.126. It has been shown that the loss function and gradient can be calculated classically as the activation function is known. Furthermore, since the proposed QFNN is a parametrized quantum circuit, there will be an exciting work to train the QFNN in a quantum way.

\section{\label{Sec5}Conclusions And Discussions}
In this paper,  we have noted that there are various oracles to implement the same Boolean function. Even for the same oracle, the circuits are not unique. We have established two circuits (see Fig.\ref{fig:function}) to implement the Boolean function $ f $, which can be regarded as an approximation of the corresponding nonlinear function. There should be more quantum circuits to approximate nonlinear functions. Moreover, this paper has proposed a generalizable framework to implement nonlinear quantum neurons that fully realizes the combination of quantum computing and neural networks. We have presented two quantum neuron examples that can be used as a fundamental building block for QNNs. The quantum neuron examples have the advantages of processing massive or ultra-high dimensional data due to the quantum parallelism and entanglement. The quantum resources required to construct a single quantum neuron are the polynomial, in function of the input size. Both IBM Quantum Experience results and numerical simulations have illustrated the effectiveness of the proposed framework, which could map the existing classical neuron designed for classical computers to quantum circuits.
	
In addition to using the controlled-phase gates and the swap test, many other existing quantum computing models can apply to our framework to realize more kinds of quantum neurons. The most far-reaching generalization of the proposed framework is not restricted to implement nonlinear quantum neurons, it may have more applications in quantum machine learning \cite{D1,D2,D3,D4}. In the future, we will connect multiple layers of our nonlinear quantum neurons to build a feedforward deep neural network, which could be fully trained on quantum computers.

\begin{acknowledgments}
The authors acknowledge the support of IBM Quantum Experience for producing the experimental results. 
This work was supported partly by the National Key R{\&}D Program of China under Grant 2018YFA0703800, the National Natural Science Foundation of China under Grants 61873317, 61873262 and 61733018, the Guangdong Basic and Applied Basic Research Foundation under Grant 2020A1515011375, and the Youth Innovation Promotion Association of the CAS.

\end{acknowledgments}

\nocite{*}


\end{document}